\documentclass{article}
\usepackage{nips07submit_e,times}

\title{Belief Propagation and Beyond for Particle Tracking}

\author{
Michael Chertkov \\
T-13, Theoretical Division \\
Los Alamos National Lab.\\
Los Alamos, NM 87545, USA \\
\texttt{chertkov@lanl.gov} \\
\And
Lukas Kroc \\
Cornell University \\
Ithaca, NY 14850, USA \\
\texttt{kroc@cs.cornell.edu} \\
\And
Massimo Vergassola \\
Institut Pasteur \\
75724 Paris Cedex 15, France \\
\texttt{massimo@pasteur.fr} }

\usepackage{bm,amsthm,amsmath,amssymb}
\usepackage{graphicx}

\usepackage{color}
\definecolor{darkdarkgreen}{rgb}{0,0.25,0}
\definecolor{darkblue}{rgb}{0,0,0.6}
\definecolor{lightgray}{rgb}{0.95,0.95,0.95}
\usepackage[bookmarks,bookmarksopen,pdfstartview=FitH,breaklinks,
            colorlinks,linkcolor=darkblue,citecolor=darkdarkgreen,urlcolor=darkblue,
            pdfsubject={}]{hyperref}

\newtheorem{prop}{Proposition}

\theoremstyle{definition}

\theoremstyle{remark}

\newcommand{\set}[1]{\left\{#1\right\}}

\begin{document}
\maketitle

\begin{abstract} We describe a novel approach to statistical learning
  from particles tracked while moving in a random environment.
  The problem consists in inferring properties
  of the environment from recorded snapshots. We consider here the case
  of a fluid seeded with identical passive particles that diffuse and
  are advected by a flow.  Our approach rests on efficient algorithms to
  estimate the weighted number of possible matchings among particles in
  two consecutive snapshots, the partition function of the underlying
  graphical model. The partition function is then maximized over the
  model parameters, namely diffusivity and velocity gradient. A Belief
  Propagation (BP) scheme is the backbone of our algorithm, providing
  accurate results for the flow parameters we want to learn. The BP
  estimate is additionally improved by incorporating
  Loop Series (LS) contributions. For the weighted matching problem, LS
  is compactly expressed as a Cauchy integral, accurately estimated by a
  saddle point approximation. Numerical experiments show that the
  quality of our improved BP algorithm is comparable to the one of a
  fully polynomial randomized approximation scheme, based on the Markov
  Chain Monte Carlo (MCMC) method, while the BP-based scheme
  is substantially faster than the MCMC scheme.
\end{abstract}

\section{Introduction}
\label{sec:Intro}

Graphical model approaches to statistical learning and inference are
widespread in many fields of science, ranging from machine learning to
bioinformatics, statistical physics and error-correction. Such
applications often require evaluation of a weighted sum over an
exponentially large number of configurations --- a formidable
$\#P$-hard problem in the majority of cases.

In this paper we focus on one such difficult problem, which occurs
when tracking identical particles moving in a random environment. As
long as particles are sufficiently dilute, their tracking in two
consecutive frames is rather straightforward. When the density of
particles and/or the acquisition time increase, many possible sets of
trajectories become statistically compatible with the acquired data
and multiple matchings of the particles in two consecutive snapshots
are likely. Despite of these uncertainties, one expects that reliable
estimates of the properties of the environment should still be
possible if the number $N$ of tracked particles is sufficiently large.
This is the problem that we want to address here.

The nature of the moving particles and their environment are not subject
to particular restrictions, e.g. they might move actively, such as
living organisms, or passively. Here, we shall consider the case of a
fluid seeded with passive particles, a problem arising in the context of
fluid mechanics experiments. Given a statistical model of the fluid flow
with unknown parameters, along with the positions of $N$
indistinguishable particles in two subsequent snapshots, one aims at
predicting the most probable values of the model parameters. This task
is formally stated in Section~\ref{sec:problem} as searching for the
maximum of a weighted sum over all possible matchings between particles
in the two snapshots. The problem turns out to be equivalent to
computing the permanent of a non-negative matrix, known to be a
$\#P$-complete problem~\cite{V:perm_comp}. The main contribution of this
paper is {\em an efficient and accurate algorithm of Belief  
Propagation (BP) type for calculating the permanent} for the class of
weight matrices arising from the particle tracking problem. The BP
algorithm seeks a minimum of the Bethe Free
Energy~\cite{YFW:constructing} for a suitable graphical model. The
graphical model is a fully connected bipartite graph: nodes are
associated with the measured particles, edges are weighted according to
the model of the flow transporting the particles and constraints enforce
the condition that exactly one edge per node is active. It is known
that BP gives the exact result for the maximum likelihood version of the
problem (finding a maximum weight matching) in spite of multiple loops
characterizing the graphical model~\cite{08BSS}. The BP algorithm for
the matching problem is derived and discussed in Section~\ref{sec:bp}.

BP equations could be understood as a re-parametrization, or gauge
transformation, of factor functions in the graphical
model~\cite{03WJW}. Furthermore, BP solutions also provide an explicit
representation of the exact partition function in terms of the
so-called Loop Series~\cite{06CCa,06CCb}. Our main technical result is
{\em the derivation of a compact expression and efficient
  approximation for the Loop Series in the problem of weighted
  particle matching}. This is done in Section~\ref{sec:loops}, where
the Loop Series is expressed in terms of an $2N$-th order mixed
derivative of an explicit functional, reduced to $2N$-dimensional
Cauchy integral and finally estimated by a saddle-point
approximation. Section~\ref{sec:experiments} describes empirical
results demonstrating the performance of bare BP and the saddle-point
improved BP in comparison with a (simplified) fully polynomial
randomized approximation scheme for computing the
permanent~\cite{JSV:poly_perm}. Our improved BP achieves comparable
accuracy, with significant gains in terms of speed. As
the number of particles tracked in experiments is typically large
(order tens of thousands) we argue that our approach is both useful
and promising for applications.

\section{Particle tracking problem}
\label{sec:problem}

An important part of modern experiments in fluid mechanics is based on
tracking of pre-seeded particles by sophisticated optical
methods~\cite{91Adr}.  If particles are sufficiently small and chosen of
appropriate (mass) density, their effect on the flow is essentially
negligible and one can safely assume that they are passively transported
by the flow. The (number) density of particles is usually rather high
and a single snapshot typically contains a large number of them. The
reason is that the smallest scales of the flow, which is generally
turbulent, ought to be resolved.  Two decades in a three dimensional
flow require to follow at least one million, $10^{2\times   3}$,
particles.  Furthermore, turbulence is quite effective in rapidly
transporting particles so that the acquisition time between consecutive
snapshots should be kept small. Modern cameras have impressive
resolutions, in the order of tens of thousands frames per second, yet
the flow of information is huge: $\sim Gigabit/s$ to monitor a
two-dimensional slice of a $(10cm)^3$ experimental cell with a pixel
size of $0.1mm$ and exposition time of $1ms$. This extremely high rate
makes it impossible to process data on the fly, unless very efficient
algorithms are developed.

Previous points motivate the development of a novel set of algorithmic
tools for fast and efficient particle tracking. One key element is
incorporating statistical models of the environment where particles
are transported and tracked. For turbulent flows, modeling proceeds as
follows. Consider $N$ particles from the same time frame, labeled by
$i=1,\cdots,N$ and positioned at the set of points $x_i$, such that
the typical distance between neighboring particles is smaller then the
viscous scale of the flow. Then, Lagrangian particles evolve according
to the set of stochastic equations, $\dot{\rho}_i=U+S\rho_i+\xi_i$,
where $\rho_i$ are particle displacements on a line (generalization to
multiple dimensions is straightforward) measured with respect to a
reference point; $U$ and $S$ are the large-scale mean and gradient of
the velocity field; $\xi_i(t)$ is the stochastic zero-mean Gaussian
Langevin noise, describing molecular diffusivity, defined by its
correlation function: $\langle\xi_i(t_1)\xi_j(t_2)\rangle=\kappa
\delta_{ij}\delta(t_1-t_2)$. Particles are indistinguishable and the
matching problem consists in assigning each particle from the original
frame $x_i=\rho_i(0)$ to particles in the subsequent frame
$y^i=\rho_i(\Delta)$. Even if the flow parameters, $U$ and $S$, and
the diffusion coefficient, $\kappa$, were known and frozen in time
(the latter is a reasonable assumption provided the acquisition time
$\Delta$ is sufficiently small), the matching cannot be identified
with absolute certainty due to the stochastic nature of diffusion. The
problem can be statistically modeled considering all possible particle
matchings $\vec\sigma$ between two successive frames and weighting
them according to
\begin{eqnarray}
  p(\vec{\sigma})=F(\vec\sigma)\cdot \prod_{(i,j)}p_i^j
  ,\quad
  p_i^j=\frac{\exp\left(-\frac{S\sigma_i^j\left(y^j-e^{S}x_i\right)^2}
      {\kappa(\exp(2S)-1)}\right)}{\sqrt{\pi(e^{2S}-1)/S}},
  \quad Z=\sum_{\vec{\sigma}}p(\vec{\sigma})\,.
 \label{prob}
\end{eqnarray}
Here, $\sigma_i^j\in\set{0,1}$ is a Boolean variable indicating
absence/presence of matching between $x_i$ and $y^j$, the vector
$\vec{\sigma}=(\sigma_i^j|i,j=1,\cdots,N)$ and
$F(\vec\sigma)=\prod_j\delta(\sum_i\sigma_i^j,1)
\prod_i\delta(\sum_j\sigma_i^j,1)$ enforces the constraints for a
perfect matching (all particles match with exactly one particle in the
other frame). For simplicity, $U=0$ (the drift common to all particles
is subtracted) and time is rescaled to have $\Delta=1$. The
partition function $Z$ is the weighted sum over all possible matchings
and $p(\vec{\sigma})/Z$ is their normalized probability
distribution. By construction, the partition function is the permanent
of the $N\times N$ positive matrix, $\hat{p}=(p_i^j|i,j=1,\cdots,N)$,
i.e. $Z=\mathrm{per}(\hat{p})$.

Our goals are: (1) For given parameters $S,\kappa$ and the set of
particle positions $\vec x$ and $\vec y$ in two subsequent frames,
have an algorithm for finding (a) the most probable matching, (b)
marginal matching probabilities for any two particles from different
frames, which is equivalent to computing the partition function. (2)
Learn and provide reliable estimates of the model parameters
$S,\kappa$.

Problem (1a) is solved by the auction exact polynomial
algorithm~\cite{92Ber}. Conversely, problems (1b) and (2) belong to the
$\#P$-complete class, i.e. are likely to be exponentially complex, and
we then aim at developing an efficient and systematically improvable
heuristics. Our approach is based on the observation, made
in~\cite{08BSS}, that a BP scheme equivalent to the auction algorithm
can be formulated for (1a), in spite of the underlying fully connected
bi-partite graph with multiple loops (see also~\cite{08Che}). Notice
that problem (1a) is the Maximum-Likelihood version of (1b). We solve
the problem (2) by taking the best possible estimate for $Z$ at given
values of $S$ and $\kappa$, and then maximizing the result over these
parameters. We observed empirically that estimates based on Expectation
Maximization (EM) algorithm~\cite{DLR:em} do not ensure accurate
learning of the flow parameters $S$ and $\kappa$ in some of the
scenarios of interest, in particular the one with diffusion only.
Conversely, BP gives accurate results in terms of the position of the
maximum w.r.t the parameters, although the estimate for $Z$ is often
orders of magnitudes wrong. To further improve on this, we apply, in
Section~\ref{sec:loops}, the general BP-based Loop Calculus approach
developed in~\cite{06CCa,06CCb} to the perfect matching problem. This
significantly improves the estimates of $Z$, especially in difficult
cases when uncertainties in the matchings are significant.

To the best of our knowledge, particle tracking as a learning problem --
not to mention the algorithmic developments based on contemporary
inference methods presented below -- is novel and it was not discussed
previously (see~\cite{06OXB} for a survey of algorithms currently used
in fluid mechanics experiments).

\section{Belief propagation and Bethe free energy}
\label{sec:bp}

For a model with states $\vec{\sigma}$ having weight $p(\vec{\sigma})$
(as in (\ref{prob})), the convex functional
\begin{eqnarray}
{\cal F}\{b(\vec{\sigma})\}\equiv
\sum_{\vec{\sigma}}b(\vec{\sigma})\ln\frac{b(\vec{\sigma})}{p(\vec{\sigma})}
\label{Gibbs}
\end{eqnarray}
has a single minimum, at $b(\vec{\sigma})=p(\vec{\sigma})/Z$ (under the
normalization condition $\sum_{\vec{\sigma}} b(\vec{\sigma})=1$), and the
corresponding value of the functional ${\cal F}$ is the free energy, $-\ln Z$.  As
shown in \cite{YFW:constructing}, the Bethe free energy approximation and BP
equations stem from (\ref{Gibbs}) by considering an ansatz of the form
\begin{eqnarray}
 b(\vec{\sigma})\approx
 \frac{\prod_i b_i(\vec{\sigma}_i)\prod_j b^j(\vec{\sigma}^j)}{
 \prod_{(i,j)} b_i^j(\sigma_i^j)}\,.
\label{BP_Belief}
\end{eqnarray}
Vectors $\vec{\sigma}_i\equiv\{\sigma_i^j|j=1,\cdots,N\}$ and
$\vec{\sigma}^j\equiv\{\sigma_i^j|i=1,\cdots,N\}$ are allowed to take any of the $N$
possible values $(0,\cdots,0,1,0,\cdots,0)$ with exactly one nonzero entry. Beliefs
$b_i^j(\sigma_i^j),b_i(\vec{\sigma}_i),b^j(\vec{\sigma}^j)$ satisfy for any $i$ and
$j$ the consistency relation for marginal probabilities: $b_i^j(\sigma_i^j)=
\sum_{\vec{\sigma}_i\setminus\sigma_i^j}b_i(\vec{\sigma}_i)=
\sum_{\vec{\sigma}^j\setminus\sigma_i^j}b^j(\vec{\sigma}^j)$, where
$\sum_{\vec{\sigma}_i\setminus\sigma_i^j}$ denotes the sum over all possible values
of the vector $\vec{\sigma}_i$ keeping fixed the value of the component
$\sigma_i^j$.  Eq.~(\ref{BP_Belief}) is exact for a tree and serves as an
approximation for graphs with loops, e.g. for the fully connected bi-partite graph
of our matching problem. Beliefs, as approximations for probabilities, should also
satisfy the normalization conditions: $ \forall (i,j), \quad b_i^j(1)+b_i^j(0)=1$.
Using the normalization and consistency conditions, we can express all beliefs via
$\beta_i^j\equiv b_i^j(1)$ and obtain for the Bethe free energy and the
normalization conditions
\begin{eqnarray}
  && {\cal F}_{BP}\{\beta\}=\sum_{(i,j)}\left(\beta_i^j\ln
\frac{\beta_i^j}{p_i^j}-
      (1-\beta_i^j)\ln(1-\beta_i^j)\right),\label{FE}\\
  && \forall i:\quad \sum_j \beta_i^j=1;\qquad \forall j:\quad \sum_i
  \beta_i^j=1. \label{short_cond}
\end{eqnarray}

A simple argument shows that constrained minima of (\ref{FE}) are
either a perfect matching (beliefs are all zeros and $N$ of them are
unity) or they are attained in the interior of the domain. The latter is the
case generally encountered in the situations of interest to us, i.e.
where no statistically dominant matching is present. Minima in the
interior are stationary points of ${\cal F}_{BP}$,
corresponding to the following set of equations:
\begin{equation}
 \forall (i,j):\quad
 \beta_i^j(1-\beta_i^j)=p_i^j\exp\left(\mu_i+\mu^j\right),\label{BP1}
\end{equation}
where the $2N$ Lagrangian multipliers $\mu$ (chemical potentials) are determined by
Eqs.~(\ref{short_cond}). Note that the Bethe free energy is not convex, and thus
multiple minima in the interior of the domain might be possible. Empirically, we
never found more than one though. In the limit where
only the Maximum Likelihood configuration is of interest, entropy terms are
discarded and (\ref{FE}) reduces to Linear Programming, yielding optimal integer
solution in accordance with \cite{08BSS,08Che}. Another relevant remark is that
convexity is restored for a modified expression of the free energy, where the minus
sign of the second term in Eq.~(\ref{FE}) is reversed. The latter expression follows
from an integral representation for $Z$, approximated in a saddle-point way. This
approximation overestimates the diffusion coefficient and we use its unique solution
(easy to find numerically) as initial condition to the following iterative version
of Eqs.~(\ref{short_cond},\ref{BP1}):
\begin{eqnarray}
 && \hspace{-1cm}\forall (i,j):\ \ \beta_i^j(n\!+\!1)\!=\!\lambda\beta_i^j(n)\!+\!
 \frac{(1-\lambda)p_i^j}{p_i^j+(\sum_k\beta_k^j(n)/2+\sum_k
\beta_i^k(n)/2-\beta_i^j(n))^2/(u_i(n)v^j(n))},
 \label{beta_n}\\
 && \forall i:\ \
 u_i(n+1)=\frac{1-\sum_j(\beta_i^j(n))^2}{\sum_k p_i^k v^k(n)},\quad
\forall j:\ \ v^j(n+1)=\frac{1-\sum_i(\beta_i^j(n))^2}{\sum_k p_k^j u_i(n)},
 \label{Vn}
\end{eqnarray}
where the arguments of the $\beta$'s indicate the order of the iterations,
$u_i=\exp(\mu_i)$ and $v^j=\exp(\mu^j)$.  The damping parameter $\lambda$ (typically
chosen $0.4\div0.5$) helps with convergence. To ensure appropriate accuracy for
solutions with $\beta$'s close to zero or unity we also insert a normalization step
after Eqs.~(\ref{beta_n}) but prior to Eqs.~(\ref{Vn}), making the following two
transformations consequently, (a) $\forall (i,j)$: $\beta_i^j\to
\beta_i^j/\sum_k\beta_i^k$, and (b) $\forall (i,j)$: $\beta_i^j\to
\beta_i^j/\sum_k\beta_k^j$.  Numerical experiments show that this procedure
converges to a stationary point of the Bethe free energy (\ref{FE}).

\section{Loop series, Cauchy integral and saddle-point approximation}

\label{sec:loops}

As shown in \cite{06CCa,06CCb}, the exact partition function of a
generic graphical model can be expressed in terms of a Loop Series
(LS), where each term of the series is computed explicitly using the
BP solution.  Adapting this general result to the matching problem,
bulky yet straightforward algebra leads to the following exact
expression for the partition function $Z$ defined in Eq.~(\ref{prob}):
 \begin{equation}
   Z=Z_{\cal BP}*z,\quad z\equiv 1+\sum_C r_C,\quad r_C=\left(\prod_{i\in C}
     (1-q_i)\right)
   \left(\prod_{j\in C} (1-q^j)\right)\prod_{(i,j)\in C}
   \frac{\beta_i^j}{1-\beta_i^j}\,.
 \label{rC}
  \end{equation}
  Here, the Bethe free energy ${\cal F}_{\cal BP}=-\ln Z_{\cal BP}$,
  the variables $\beta$ are in accordance with
  Eqs.~(\ref{short_cond},\ref{BP1}), and $C$ stands for an arbitrary
  generalized loop, defined as a subgraph of the fully connected
  bi-partite graph with all its vertexes having degree of connectivity
  $>1$. The $q_i$ (or $q^j$) in Eq.~(\ref{rC}) is the $C$-dependent
  degree of connectivity of nodes, i.e. $q_i=\sum_{\set{j \mid
      (i,j)\in C}} 1$ and $q^j=\sum_{\set{i \mid (i,j)\in C}}
  1$. According to Eq.~(\ref{rC}), loops with even/odd number of
  vertexes give positive/negative contributions $r_C$. Therefore, the
  series is not positive definite, which is also consistent with the
  fact that $Z_{\cal BP}$ in general does not provide a lower bound
  for the exact partition function. (In some special cases, e.g. for
  the model studied in \cite{07SWW}, all terms in the series are known
  to be positive and thus $Z_{\it BP}\leq Z$.)  In all cases of the
  weighted matching problem we have experimented with, we have
  empirically found that the inequality $Z_{\it BP}<Z$ still
  holds. Let us finally notice an important special feature of the
  weighted matching problem: for any generalized loop $C$, its
  individual contribution $|r_C|\leq 1$. The proof can be found in 
  the Appendix.

  Eq.~(\ref{rC}) allows for the following compact representation in
  terms of $2N$-th order mixed local derivative of an explicit
  function of $2N$ variables
\begin{eqnarray}
 && z=\left.\frac{\partial^{2N} {\cal
 Z}(\rho_1,\cdots,\rho_N,\rho^1,\cdots,\rho^N)}
 {\partial\rho_1\cdots\partial\rho_N\partial\rho^1\cdots\partial\rho^N}
 \right|_{\rho_1=\cdots=\rho_N=\rho^1\cdots=\rho^N=0} ,
 \label{ZN1}\\
 && {\cal Z}(\vec{\rho})\equiv \exp\left(\sum_i\rho_i+\sum_j\rho^j\right)
 \prod_{(i,j)}
 \left(1+\frac{\beta_i^j}{(1-\beta_i^j)}\exp\left(-\rho_i-\rho^j\right)\right),
 \label{ZN2}
\end{eqnarray}
where $\vec{\rho}=(\rho_1,\cdots,\rho_N,\rho^1,\cdots,\rho^N)$ are
auxiliary variables. However, calculating the $2N$-order mixed
derivative exactly is a task of exponential complexity and one wonders
whether the mixed derivative can be approximated efficiently. Partial
answer to this question is given below.

Using the Cauchy integral representation for the first-order derivative of an
analytic function (${\cal Z}$ is analytic over $\rho_i,\rho^j$ with finite real
parts), Eqs.~(\ref{ZN1},\ref{ZN2}) can be recast as the following contour integral:
\begin{eqnarray}
  && z=\oint_{\Gamma_\rho}\exp\left(-{\cal G}(\vec{\rho})\right)\frac{\prod_i
    d\rho_i
    \prod_j d\rho^j}{(2\pi i )^{2N}},\quad {\cal G}(\vec\rho)\equiv \sum_i
  2\ln\rho_i+\sum_j 2\ln\rho^j-\ln{\cal Z}, \label{ZN3a}
\end{eqnarray}
where $\Gamma_\rho$ is a direct product of $2N$ close contours
circling clockwise the origin $\vec{\rho}=\vec{0}$, where derivatives
in (\ref{ZN1}) are to be computed. We observe that each integral over
an individual $\rho$ variable in Eq.~(\ref{ZN3a}), say $\rho_i$, has a
pole at $\rho_i=0$ and essential singularities at
$\rho_i=\pm\infty$. Notice also that ${\cal G}(\vec{\rho})$ is a
concave function of $\vec{\rho}$ in each one of the $2^{2N}$ quadrants,
i.e.  where components of $\vec{\rho}$ are finite, real and have a
definite sign. Therefore, it is natural to shift the contour of
integration to one of $2^{2N}$ maxima of ${\cal G}(\vec{\rho})$,
\begin{equation}
  \forall i:\
  \frac{2}{\varrho_i}=1-\sum_j\left(1+\frac{1-\beta_i^j}{\beta_i^j}
    e^{\varrho_i+\varrho^j}\right)^{-1},
  \quad \forall j:
  \frac{2}{\varrho^j}=1-\sum_i\left(1+\frac{1-\beta_i^j}{\beta_i^j}
    e^{\varrho_i+\varrho^j}\right)^{-1},
 \label{ZN5}
\end{equation}
and orient the contour along the direction of steepest descent from
the saddle point. Once the sign of each component of $\vec{\rho}$ is
fixed, approximating the respective solution of Eqs.~(\ref{ZN5}) numerically
is straightforward due to the concavity of ${\cal G}$. We shall
enumerate the various maxima by the index $s$.

The saddle-point approximation of Eq.~(\ref{ZN3a}), accounting for Gaussian
integral corrections about all the maxima of Eq.~(\ref{ZN5}), yields
\begin{equation}
 z\approx\sum_{s=1}^{2^{2N}}
\frac{\exp\left(-{\cal G}(\vec{\varrho}^{(s)})\right)}
 {(2\pi)^N\sqrt{\det(\hat{\Lambda}^{(s)})}}\equiv \sum_{s=1}^{2^{2N}}
 \exp\left(-{\cal G}^{(s)}_{sp}(\vec{\varrho}^{(s)})\right) ,\label{sp1}
 \end{equation}
 where $\hat{\Lambda}^{(s)}$ is the Hessian of ${\cal G}(\vec{\rho})$
 at the saddle-points. Corrections to each term in Eq.~(\ref{sp1}) are
 measured in terms of higher-order terms, the leading (fourth-order)
 being estimated as $ {\cal
   G}^{(s)}_{4}=-\frac{1}{8}\sum_{\alpha,\beta,\gamma,\nu}
 \Upsilon^{(s)}_{\alpha\beta\mu\nu}
 ((\hat{\Lambda}^{(s)})^{-1})_{\alpha\beta}((\hat{\Lambda}^{(s)})^{-1})_{\mu\nu}$,
 where $\Upsilon^{(s)}_{\alpha\beta\mu\nu}$ is the tensor of
 fourth-order derivatives of ${\cal G}(\vec{\rho})$. The improved
 approximation for the partition function becomes $ z\approx\sum_s
 \exp(-{\cal G}_{sp}(\vec{\varrho}^{(s)})-{\cal
   G}_4(\vec{\varrho}^{(s)}))$. One reason to account for the
 fourth-order term is that the ratio $|{\cal G}_4/{\cal G}_{sp}|$
 gives a standard measure of the saddle-point validity. In cases where
 the saddle-approximation becomes asymptotically exact, the ratio
 approaches zero. A weaker condition, $|{\cal G}_4/{\cal G}_{sp}|<1$,
 suffices for a heuristically reasonable approximation.

 We also expect that the typical order of magnitude of the $2^{2N}$
 terms ${\cal G}_{\it sp}^{(s)}$ and ${\cal G}_4^{(s)}$ grows as
 $O(N)$. Therefore, the sum in $s$ over the saddle-points might be
 dominated by a single term in the limit of large $N$.  As
 discussed in the next Section, we find empirically that such dominant
 term indeed exists and happens to correspond to the vector
 $\vec{\varrho}^{(+)}$ having all its components positive.  Therefore,
 the sum over all the maxima, indexed by $s$, can be simplified by
 keeping only the dominant contribution correspondent to $\vec{\varrho}^{(+)}$. The
 expression that we have employed in numerical experiments discussed
 in the next Section reads finally,
 $z\approx \exp\left(-{\cal G}_{sp}(\vec{\varrho}^{(+)})-{\cal
 G}_4(\vec{\varrho}^{(+)})\right)$.

\section{Numerical results}
\label{sec:experiments}

In this Section we compare the accuracy of BP based approximations for
the partition function, $Z$, of the model~(\ref{prob}) with MCMC
simulations. As briefly stated in Section~\ref{sec:problem}, computing
$Z$ for the weighted matching problem is equivalent to computing a
permanent of a non-negative matrix, for which a Fully Polynomial
Randomized Approximation Scheme (FPRAS) exists based on the Markov
Chain Monte Carlo method~\cite{JSV:poly_perm}. We implemented the
basic idea of FPRAS, with some simplifications applicable to our
problem. This algorithm was used to assess accuracy of our
approximations, but is orders of magnitude slower than the BP
based approaches, and thus not applicable to large particle tracking
problems.

To study dependence of $Z$ on $\kappa$ and $S$ (see discussion of
Section~\ref{sec:problem}), we estimate $Z$ at different values of
$\kappa$ and $S$ and compare the curves, searching for the maximum
with respect to these parameters. Our BP simulations consist of the
following steps for each value of $\kappa$ or $S$. First, we find
solution of BP equations running the numerical scheme described in
Eqs.~(\ref{beta_n},\ref{Vn}) and calculating the resulting ${\cal
  F}_{BP}$ according to Eq.~(\ref{FE}) . Second, we find the solution of
the saddle-point Eqs.~(\ref{ZN5}) in the various quadrants, calculate
the covariance matrix $\hat{\Lambda}^{(s)}$ for this saddle solution
and thus estimate the leading saddle-point correction ${\cal G}_{sp}$
in accordance with Eq.~(\ref{sp1}). Finally, we estimate the
respective fourth-order correction, ${\cal G}_4^{(s)}$.

We compare respective contributions to the partition function
associated with saddle-points with different choices of the signs.  In
all experiments where the typical overlap among particles is
significantly smaller than the total number of particles we find that
the contribution with all signs $+$ dominates. Moreover, the gap
separating the leading $+$ contribution and other contributions is
significant and grows linearly with $N$. This allows us to ignore all
other saddle-point contributions but the $+$ one. (In cases of
moderate $N$ we have made the exhaustive comparison.  In general, we
compared the all $+$ contribution with all $-$ contribution,
contributions with a limited number of signs flipped and also with
signs generated randomly.) We observe that the saddle-point validity
conditions holds reasonably well if $N$ is sufficiently large, at
$N=100$ the ratio $|{\cal G}_4/{\cal G}_{sp}|$ is typically $0.1\div
0.4$. Since $|{\cal G}_4/{\cal G}_{sp}|\to 0$ when $S\to +\infty$,
we find that the saddle-point is very accurate
(possibly asymptotically exact) in this limit.

\begin{centering}
\begin{figure}
\includegraphics[width=2.8in]{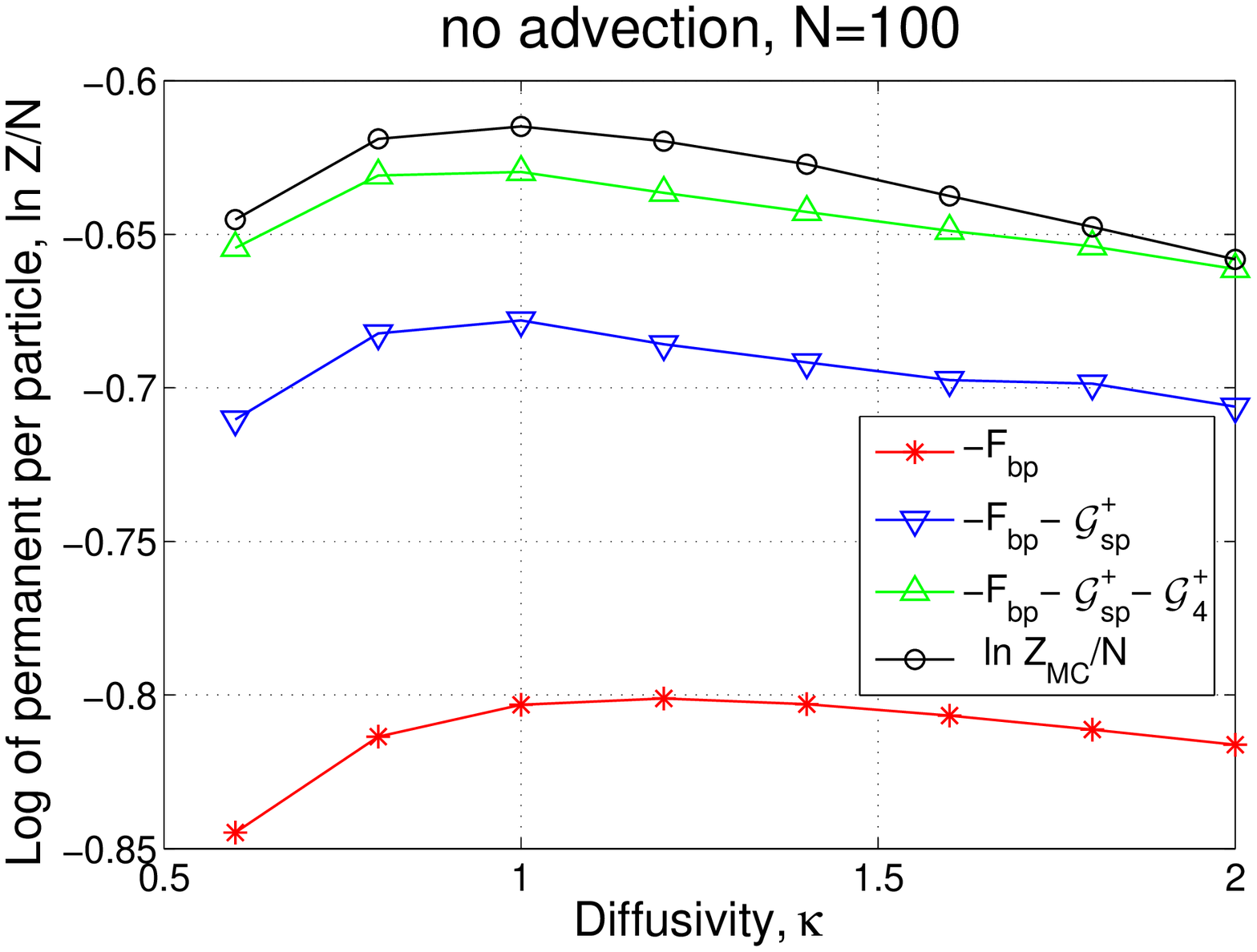} \hfill
\includegraphics[width=2.8in]{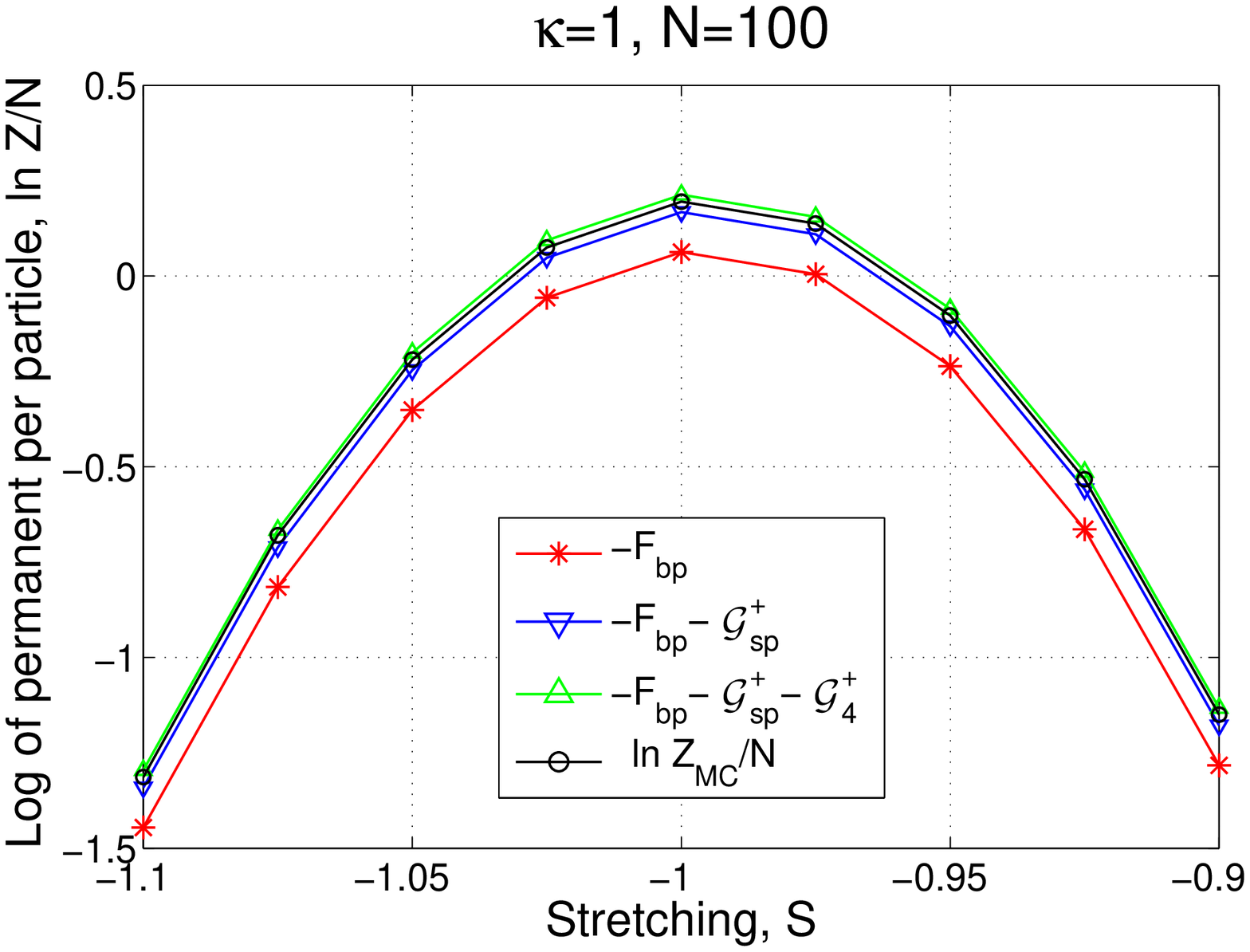}
\caption{Numerical results comparing BP, Loops series improvement and MCMC. Left
panel: diffusion only; right panel: diffusion and advection. Red curve are BP
results, blue and green are Loop series improvements, and black curve is MCMC.}
\label{fig:exps}
\end{figure}
\end{centering}

Results of numerical simulations for number of particles $N=100$ are
shown in Figure~\ref{fig:exps}. The plots show results for one set of
particle positions each, but we found that variability between scenarios
decreases with increasing number of particles, and one scenario is thus
sufficient to characterize the main trends.

The left panel shows results for a situation with diffusion only, with actual
diffusivity $\kappa_{\mbox{act}}=1.0$ used for generating the particle positions.
The x-axis spans values of the governing parameter $\kappa$ (pretended not to be
known), and the y-axis shows configuration weight per particle (what is shown is
$\ln Z/N$).  The black curve are results of the MCMC simulation, which are close to
the true values of $Z$.  The curve indeed peaks around the correct value
$\kappa=\kappa_{\mbox{act}}=1.0$, although the maximum is rather shallow (the lower
$\kappa_{\mbox{act}}$ the more pronounced is the maximum). The red curve corresponds
to results obtained by BP only, and shows that BP severely underestimates the
partition function, although its maximum is at the right value. The remaining blue
and green curves are the two saddle corrections discussed in Section~\ref{sec:loops}
\footnote{We find   that the quality of saddle-approximation decreases with
$\kappa$, when the number of ``polarized'' matchings, i.e. those with $\beta_i^j\to
1$, increases. Polarized beliefs, correspondent to almost committed matchings, do
not contribute significantly to the Loop Series yet their respective contributions
are misrepresented in the saddle-approximation.  To compensate for this caveat of
the saddle-approximation we decimated the original graph, reducing it to   a
subgraph with all particles involved in ``polarized'' matchings (and all edges
associated with them) pruned out.  For the numerical experiments in
Figure~\ref{fig:exps}, we used the polarization criterion, $\beta_i^j>0.01$.}.  As
can be seen, the saddle corrections significantly improve the BP estimate.  The time
to compute each point in the plot using the BP based scheme is $\approx 5 sec $,
while the MCMC algorithm takes $\approx 10 mins$.

The right panel of Figure~\ref{fig:exps} shows results for a scenario
with both advection and diffusion. The x-axis now spans values of the
velocity gradient $S$, and y-axis is again $\ln Z/N$. The actual
velocity gradient used for generating the particles was
$S_{\mbox{act}}=-1.0$. The four curves show similar main trends: BP
underestimates $Z$ and loop corrections provide very tight belt around
the MCMC results. The main difference is that in the case when advection
is present, the peak is very well pronounced and all methods give
extremely accurate answer for the velocity gradient $S$. In this case,
the running times were again $\approx 5 sec$ for the BP scheme, but
$\approx 5 hours$ for MCMC per point.

\section{Conclusions}

We have presented new computational tools for particle tracking, based on Belief
Propagation and Loop Series and compared them to the Markov Chain Monte Carlo scheme
for the estimation of the permanent~\cite{JSV:poly_perm}. We have specifically
considered tracking of passive particles in fluid dynamics experiments. The methods
are quite general though and applications to other tracking problems, e.g. to
self-propelling biological objects, are possible and will be pursued. Our long-term
goal is to develop computational tools effective enough to permit on-the-fly
processing of particle tracking frames. Algorithms presented here ensure an
excellent accuracy and the BP-improved scheme is already orders of magnitudes faster
than the MCMC scheme.

\appendix

\section*{Appendix}
Here we prove the following important property of the loop series for perfect
matching:

\begin{prop}
In the loop calculus expansion of the graphical model for perfect
matching problem~(\ref{rC}), $|r_C| \leq 1$
for all generalized loops $C$.
\end{prop}
\begin{proof}
We rewrite expression~(\ref{rC}) as: $r_C = \left(\prod_{i\in C}
\psi_{i;C}\right)\left(\prod_{j\in C} \psi^j_{C}\right)$, where $\psi_{i;C} =
(1-q_i)\prod_{\set{j \mid (i,j)\in C}}\sqrt{\beta_i^j/(1-\beta_i^j)}$, and
analogously for $\psi^j_C$. We will use the fact that in a fixed point of our BP
scheme $\sum_i \beta_i^j = \sum_j \beta_i^j = 1$. We proceed by showing that
$|\psi_{i;C}|$ is maximized for $\beta_i^{j_1}=\ldots=\beta_i^{j_{q_i}}=1/q_i$, and
that even for such beliefs $|\psi_{i;C}|\leq 1$. The situation is analogous for
$\psi_C^j$. It then follows that $|r_C| \le 1$ as desired.

Let us fix a generalized loop $C$ and a node $i$ with degree $q_i\geq 2$ in $C$ and
seek values of
$\beta_i^{j_1}=\ldots=\beta_i^{j_{q_i}}\in[0,1]\mbox{~s.t.~}\sum_{k=1}^{q_i}
\beta_i^{j_k}\leq 1$ that maximize $|\psi_{i;C}|$. First, by observing that
$\beta/(1-\beta)$ is a growing function of $\beta$, we see that indeed
$\sum_{k=1}^{q_i} \beta_i^{j_k}=1$ must be the case. Focusing only on the
subexpression of $\psi_{i,C}$ with $\beta$ in it, we have the following constrained
optimization problem: $\max_\beta \prod_{j:(i,j)\in C}
\frac{\beta_i^j}{1-\beta_i^j}\mbox{~s.t.~}\sum_{k=1}^{q_i} \beta_i^{j_k}=1$. We
apply the method of Lagrangian Multipliers, and arrive at the following conditions
for extrema:
\[
\forall i,j:~ \frac{\partial}{\partial \beta_i^j} \left( \prod_{\set{j \mid (i,j)\in
C}} \frac{\beta_i^j}{1-\beta_i^j} + \lambda\left(1-\sum_{\set{j \mid (i,j)\in
C}}\beta_i^j\right)\right)= \frac{1}{(1-\beta_i^j)^2} \prod_{j'\neq
j}\frac{\beta_i^{j'}}{1-\beta_i^{j'}} - \lambda = 0
\]
Taking any two $j_1 \neq j_2$, we derive
$\beta_i^{j_2}(1-\beta_i^{j_2}) = \beta_i^{j_1}(1-\beta_i^{j_1})$ from the
above.
This, being a quadratic equation in $\beta_i^{j_1}$, has exactly two
solutions, and it is not difficult so see that they must be
$\beta_i^{j_1}=\beta_i^{j_2}$ or $\beta_i^{j_1}=1-\beta_i^{j_2}$. These
are the conditions on $\beta$s for an extremum of the product.

To incorporate the $(1-q_i)$ subexpression of $\psi_{i;C}$, let us deal with the
special case when $q_i=2$ separately: in this case, since
$\beta_i^{j_1}+\beta_i^{j_2}=1$, we have that the whole product we maximize is equal
to $1$, and thus $|\psi_{i;C}|=1$. For $q_i>2$: if, for {\em any} pair $j_1, j_2$:
$\beta_i^{j_1}=1-\beta_i^{j_2}$, then all other $\beta$s (other than $\beta_i^{j_1}$
and $\beta_i^{j_2}$) must be $0$ due to $\sum_{k=1}^{q_i} \beta_i^{j_k}=1$. This
means that the product is zero, and therefore $\psi_{i;C}=0$. If, on the other hand,
$\beta_i^{j_1}=\beta_i^{j_2}$ for all pairs $j_1, j_2$, we have that
$\beta_i^j=1/q_i$. In this situation, the question is whether $(q_i-1)^{1-q_i/2}\leq
1$, which is true for any $q_i\geq 2$. In fact, we have shown that the expression
$(q_i-1)^{1-q_i/2}$ is an upper bound on $|\psi_{i;C}|$ for any node $i$ in any
generalized loop $C$.

\end{proof}

\bibliographystyle{plain}
\small{
\bibliography{bpMatching}
}

\end{document}